# The future of astronomy with small satellites

Stephen Serjeant, Martin Elvis, Giovanna Tinetti

## Abstract

The number of small satellites has grown dramatically in the past decade from tens of satellites per year in the mid-2010s to a projection of tens of thousands in orbit by the mid-2020s. This presents both problems and opportunities for observational astronomy. Small satellites offer complementary cost-effective capabilities to both ground-based astronomy and larger space missions. Compared to ground-based astronomy, these advantages are not just in the accessibility of wavelength ranges where the Earth's atmosphere is opaque, but also in stable, high precision photometry, long-term monitoring and improved areal coverage. Astronomy has a long history of new observational parameter spaces leading to major discoveries. Here we discuss the potential for small satellites to explore new parameter spaces in astrophysics, drawing on examples from current and proposed missions, and spanning a wide range of science goals from binary stars, exoplanets and solar system science to the early Universe and fundamental physics.

## Main

Astronomers tend towards wanting ever larger telescopes to collect more light, so the utility of small satellites is not always intuitive. Compared to ground facilities, space has many advantages. In the past the main deterrent to build more space instrumentation has been the prohibitive cost. However, over the past two decades the cost per kg of launches to low Earth orbit (LEO) has been dramatically reduced by factors of several with the entry of commercial operators (e.g.[1]). Meanwhile, assembly lines offer the prospect of reducing costs associated with satellite design and manufacture for a fleet of identical or near-identical spacecraft. These developments create new opportunities for astronomers to exploit with creative approaches, despite their limitations of scale.

The economic demand for internet bandwidth has led to commercial operators entering this domain, such as OneWeb and SpaceX. Can these developments also drive new mission concepts in astronomy and fundamental physics? This Perspective will mainly focus on 400-1500 km orbits, low masses (e.g. <50 kg), small physical sizes (e.g. 700x600x200 mm), and long lifetimes (e.g. 5 years); however, to allow for technical developments, some of the concepts may currently exceed one or more of these envelopes. This article follows in part from a symposium on this topic held at the Nature offices in London on 11 November 2019.

This article would not be complete without at least a brief mention of the challenges posed to astronomy by constellations of these same small satellites. Tens of thousands of objects in LEO (e.g. 550 km) are expected to cause satellite streaks in most twilight observations by the Rubin Observatory, while tens of thousands of satellites in higher orbits of ~1100 km would be far more problematic[2]. Such bright tracks can cause complex cross-talk effects across the images. Nor is this problem restricted to wide-field optical astronomy. The Square Kilometre Array (SKA) is being built at sites protected by national legislation from ground-based radio signals, but only part of the 8.3-15.3 GHz "Band 5b" window is protected for astronomy so observations in this band are still susceptible to interference from satellite downlink signals. Investigations are underway on quantifying the potential scientific and technical impacts in other wavelength domains, such as in infrared astronomy or in Cherenkov telescopes. Even space astronomy is affected, with satellite trails routinely seen in data from LEO observatories such as the Hubble Space Telescope[3], the CHEOPS exoplanet mission, and the WISE infrared telescope[4], and satellite trails are expected to be a more significant problem for e.g. the planned 2m-class wide-field telescope associated with the Chinese space station (M. McCaughrean, personal comm.). This field is moving very quickly, with active and constructive engagement between the academic and engineering communities in industry. This work includes the possibility of impact mitigations through the modifying of observing campaigns, impact mitigations through satellite design and deployment, as well as mitigation measures for the increased risks of satellite collisions. However, this takes us beyond the scope of this article.

Small satellites and satellite constellations nevertheless also present opportunities to astronomy. In this Perspective, we discuss some of these opportunities, and we touch on some of the technical challenges. Our aim is not to be comprehensive as that would not be possible in this short article, but rather to highlight examples in a few critical areas for astronomy.

# The Motivation for Space Astronomy

Space has historically been expensive to reach, so there must be strong practical and technical drivers for preferring space platforms over terrestrial ones for some scientific use cases. Indeed there are. The first driver is atmospheric opacity. This does not just mean temporary occlusion from clouds, though these are certainly a limiting factor to some ground-based observations. Astronomy is fundamentally a multi-wavelength discipline (e.g.[5,6]), and in order to build a complete physical understanding of almost all objects and processes in astronomy, usually one needs the whole electromagnetic spectrum. Fig. 1 shows the atmospheric transparency as a function of wavelength. The atmosphere is transparent in the optical, and for a large part of the radio, and fairly transparent in a few mid-infrared windows. But if the science objective requires observations in the ultraviolet (UV), or X-rays, or gamma rays, or in most of the infrared, or at very long radio wavelengths, then the only way this can be achieved is to perform the data collection above the atmosphere. One approach to mitigating the limitations of terrestrial astronomy is to work at high altitudes, where the transparencies are better, especially in the infrared (e.g.[7]). In the regions of partial transparency in the mid-infrared and submillimetre wavelengths, the opacities are dominated by absorption from water vapour, favouring dry, high-altitude observatories (Fig. 1), but even at these sites the weather fluctuations can cause transparencies to vary from tens of percent to zero[7,8]. The improved water vapour transparency at high altitude is part of the reason for launching observatories on weather balloons (e.g.[16]), and it is part of the reason why many ground-based observatories are at high altitudes. It is also one reason why a few astronomical observations are done from sounding rockets (e.g.[17]), but observations can have much longer durations in orbit than the few minutes a sounding rocket can provide.

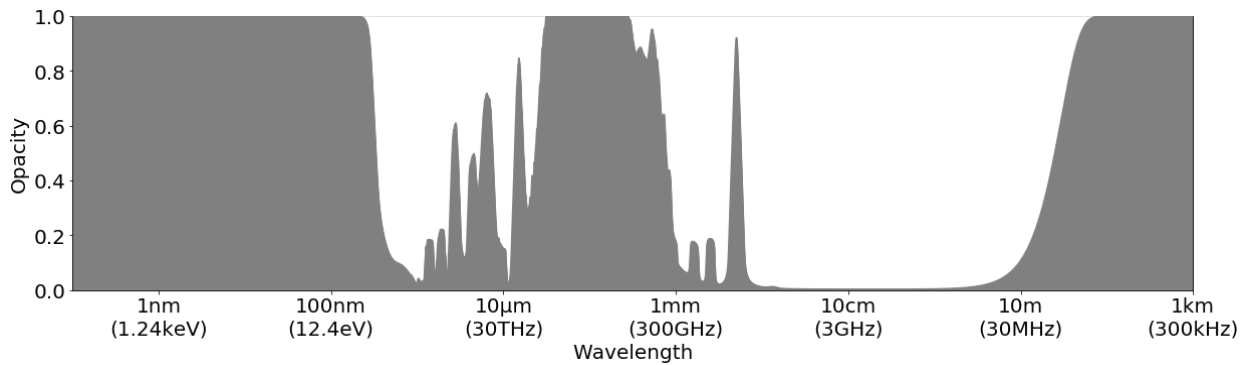

*Figure 1 Zenith opacity at the summit of Mauna Kea (altitude 4.2 km) as a function of wavelength, with selected conversions to frequency and energy. Note the transparent windows at optical (~400-700 nm) and most radio wavelengths, limited infrared transmission, and complete opacity in the X-ray and very long wavelength radio. The diagram assumes a water vapour column of 1.0 mm Hg ([7,8], Gemini Observatory) and median optical transmission ([9], Canada France Hawaii Telescope Observers' Manual). Below 310nm the optical depths are approximated by an extrapolation in log-wavelength from the near-ultraviolet cut-off, equivalent to a power-law radiative amplification factor (e.g.[10]); below ~280nm and between 28-190µm the atmosphere is essentially completely opaque even at this high-altitude observatory site[11,7,12]. The data have been smoothed with an 0.1dex boxcar kernel. The ionospheric cut-off assumes a frequency-dependent optical depth[13] of $0.01(\nu/100 MHz)^{-2}$. At GeV energies and above, photons undergo electron-positron pair production interactions in the upper atmosphere, generating extensive air showers of charged particles, whose Cherenkov light can be detected by wide-field ultraviolet and optical ground-based telescopes with very high time resolution (e.g.[14,15]).*

Another reason to go to space is atmospheric instability. The atmosphere is not just (selectively) opaque, it is also turbulent. This affects in real time the shape of the point spread function (the apparent shape of a point-like object such a star). This turbulence can be seen in real time even with some amateur telescopes trained on the Moon. For large telescopes, the distortion varies across the field of view, requiring adaptive optical technologies to measure the distortion in the incoming wavefront and correct for it in real time (e.g.[18]), either using a nearby reference star, or using lasers to fluoresce in the upper atmosphere to obtain a reference, or by "lucky imaging", i.e. imaging with high time resolution and disregarding the bulk of the data to retain only the serendipitously least affected data (e.g.[19]). Unaided, the angular resolution of ground-based optical and infrared astronomy is rarely better than 0.5 arcseconds (zenith, 500nm / 600THz) at even the best observatory sites, with image quality degrading with both airmass and frequency[18]. Lucky imaging can improve image quality by factors[19] of between 2.5 and 4 with bright guide stars (e.g. I-band magnitude 16), while adaptive optics corrections can reduce the 50% encircled energy diameter to as little as tens of milliarcseconds[18], but telescope diffraction limits are typically achievable only at wavelengths longer than visible light and only in relatively rare good seeing conditions. With space astronomy, imaging is much more stable (as well as being naturally diffraction-limited), which is critical for example for monitoring stars for subtle changes, such as exoplanets passing in front of their stars. Atmospheric stability is also not just a problem for optical astronomy. In the radio, the distortions come from the ionosphere, which changes the apparent positions of objects in real time, as well as the sizes, the phases of the incoming waves, and much else (e.g.[20]). The ionosphere is the single biggest systematic uncertainty in some radio observations (e.g.[21,13]).

A further reason to prefer space astronomy for some scientific applications is the presence of terrestrial emissions. This is particularly the case for infrared astronomy, because the Earth and its atmosphere are bright and radiate thermally with a peak around 10μm. This emission spectrum in the near-infrared has "airglow" emission lines mostly from the OH molecule, so ground-based infrared spectroscopy has additional problems. In the radio, there is the problem of terrestrial radio frequency interference (RFI, e.g.[22]).

The following sections will provide examples of scientific applications of space astronomy in small satellite contexts, with particular regard to exploiting these advantages of space platforms.

# Radio Astronomy

Radio astronomy has one of the few probes of how the "dark age" of the Universe - the time before any stars had ignited - ended. Prior to the epoch at which we see the Cosmic Microwave Background (CMB), the Universe was an opaque and fully ionised plasma. As the Universe cooled and expanded, this plasma combined into neutral gas in a process known as *recombination,* despite it being the first formation of neutral atoms in the history of the Universe. At approximately the same time the photon optical depth dropped below unity, the mean free path of photons tended to infinity, and the Universe became transparent. This light from the epoch of last scattering is the CMB[23,24]. With no further luminous sources the Universe entered its dark age until, a few hundred million years after the Big Bang, the first stars formed, and they in turn reionised their surroundings. The bubbles of ionisation (Strömgren spheres) around these stars create characteristic absorption and emission features against the CMB in the neutral hydrogen electron spin-flip transition at a wavelength of 21 cm in the radio band (e.g.[25]). These features are the signatures of the first ionising sources of light in the Universe, and the detection and characterisation of this reionisation process is one of the principal science goals of the next generation of radio facilities such as the Square Kilometre Array (SKA, e.g.[21]).

The next generation of infrared space observatories such as the *James Webb Space Telescope* (JWST, e.g.[26]), *Euclid* [27] and the *Nancy Grace Roman Space Telescope*[28] will attempt to detect the ionising sources themselves, especially when aided by magnification from strong gravitational lensing[29,30]; these will be discussed in more detail in the next section. SKA precursor and pathfinder facilities such as the Low Frequency Array (LOFAR[31]) are also attempting to characterise this reionisation signal. No detection has yet been made, and all the experiments are technically challenging (e.g.[32]). The LOFAR antennas across Europe detect signals from most of the sky, and the varying astrophysical signal across the sky is separated from ionospheric foregrounds and reconstructed in an enormous computational effort. Avoiding the ionosphere therefore has obvious attractions.

There have therefore been several proposals for small satellite swarms above the ionosphere to map the radio sky at ultra-low-frequencies[33], such as the Space-Based Ultra Low Frequency Radio Observatory (SULFRO[34]), the Dark Ages Polarimeter Pathfinder (DAPPER[35]), the Orbiting Low Frequency Antennas for Radio Astronomy (OLFAR[36]), the Distributed Aperture Array for Radio Astronomy In Space (DARIS[37]), the Formation-flying sub-Ionospheric Radio astronomy Science and Technology (FIRST[38]), the Space-based Ultra-long wavelength Radio Observatory (SURO[39]), the Discovering the Sky at the Longest Wavelengths mission (DSL[40]), Nanosatellites pour un Observatoire Interferometrique Radio dans l'Espace (NOIRE[41]), etc. In nearly all these potential dark ages experiments, the satellites are located either at the second Earth-Sun Lagrange point or in Lunar orbit. The reason for this is the RFI from the Earth. If one can eclipse the Earth by the Moon, one achieves a very radio-dark site. Low-Earth orbit would therefore present significant additional technical challenges, but the AERO/VISTA mission is a proposed microsatellite to investigate RFI mitigation[42]. The objective is to test whether one can relax the requirement that space-based low frequency interferometers must be placed far from the Earth, suggesting a tantalising possibility that dark ages radio experiments may eventually be achievable more easily and affordably in LEO.

# Infrared Astronomy

The simultaneous needs of minimising foreground emissions and having access to spectral windows (see above) have led to a long history of space applications in infrared astronomy. (Optical astronomy, where the advantage of space is usually in diffraction-limited photometric stability rather than improved transparency or backgrounds, will be discussed mainly under Exoplanets below.) Many landmark results have followed from very large (costly) missions, such as the *Infrared Astronomy Satellite* (IRAS[43]), the *Infrared Space Observatory* (ISO[44]), *AKARI* [45], *Spitzer*[46], the *Wide-field Infrared Survey Explorer* (WISE[47]), *Herschel*[48] and in the future SPHEREx[49], JWST[26] and SPICA[50]. Costs for all these missions are in excess of $100 million, with a few approaching the billion and one the 10 billion price tag. However, there are still many scientific use cases where the technical requirements can be fulfilled, in some cases uniquely, by much smaller missions. For example, the Infrared SmallSat for Cluster Evolution

Astrophysics (ISCEA[51]) seeks to map selected galaxy clusters at the peak epoch of star formation (redshifts around 2) and their surrounding cosmic webs, through near-infrared imaging and spectroscopy (avoiding the terrestrial thermal background, OH airglow and restricted atmospheric transmission), and at a cost of <$35 million without launch costs.

The radio missions in the previous section have reionisation as a principal science goal, but another complementary approach is to seek to detect the integrated redshifted light from the first stars. To detect the first individual star clusters themselves requires giant future facilities such as JWST, but the integrated light from across the whole sky is a more tractable detection problem. Because of redshifting, this light should appear at near-infrared wavelengths as a cosmic near-infrared extragalactic background (CNIRB). The spectrum of this background is strongly diagnostic of the ionising sources. Several satellite and rocket missions have sought to detect this CNIRB. The principal difficulty is that scattered sunlight from Zodiacal dust in our solar system peaks at similar wavelengths, and is several times brighter[52]. This dust is also not simply distributed throughout the Solar System but has several intricate features making the foreground signal time- and position-dependent (e.g.[53]). Zodiacal dust emission removal is the principal source of systematic errors in cosmic near-infrared background measurements, and in every CNIRB detection experiment there is considerable debate over the control of this systematic. Therefore, the CIBER[54] sounding rocket experiment sought to measure the Fraunhofer absorption lines in the extended light. These lines are present in the Sun's atmosphere, but there are no additional Fraunhofer absorption features from the dust scattering process, so one can obtain an estimate of the amount of scattered sunlight from the depths of these absorption lines. The CIBER experiment had only a 10 minute flight time, but a payload on a small satellite, or several small satellites, could have a much longer duration and could sample a much wider range of lines of sight and solar elongation angles.

# High Energy Astronomy

At first blush high energy astronomy is an unlikely regime in which to make breakthroughs with smallsats. High energy astronomy refers to radiation reaching us anywhere in the eight plus decades of photon energy (or equivalently frequency) that lie above the point at which hydrogen is ionised, the Lyman limit at 13 electron-Volts (eV, corresponding to a wavelength of 91.2 nm). Traditionally this broad band is divided into extreme ultraviolet (EUV, up to roughly 200 eV), X-ray (roughly from 2 keV (kilo-eV) to 200 keV) and gamma-ray (roughly 200 keV to 2 Giga-eV) astronomy. Across this entire band the number of photons coming from all the individual sources in the sky is rarely more than ten per second per square meter, requiring detectors larger than a smallsat can carry.

Luckily that impression misses major developments in both astrophysics and technology that open up several strong contenders for doing great science on small space platforms. Astrophysically the key point is that the sky is dynamic, particularly at high energies. Briefly high energy sources can be extremely bright, so that small detectors are sufficient. Lately it has been recognised that there are large amplitude changes occurring on virtually all timescales.

The ubiquity of these transient events has led to a dynamic new sub-field called Time Domain Astronomy. These events range from finding the gamma-ray counterparts to gravitational wave chirps, to exoplanets transiting their host stars that, at high energies, can probe the loss of a planet's atmosphere due to heating from its star, and so its habitability.

Many of these events occur at random over the whole sky at a fairly low rate, and that pushes us to cover the whole sky continuously in order to build up large enough samples to study in a reasonable time. Constellations of smallsats in LEO are the technological response to the need to cover all the sky, all the time.

There are two approaches to covering the whole sky: timing and imaging. The Hermes[55] mission makes use of 10 microsecond timing of gamma-ray bursts from a hundred or more 3U cubesats, a set of 10 cm-on-a-side unit cubes (each referred to as 1U) connected in a line of three. The precise timing of each gamma-ray burst is different on opposite sides of the Earth at this level. That difference pins down the location of the burst to 10 arcseconds, sufficient to identify the galaxy from which it originated. As [56] emphasise, this precise timing can make tests of fundamental physics too, including the quantisation of space-time (assuming the effects are first order in *v/c*) by determining microsecond differences in light travel times of billions of years. The alternative imaging approach uses a far smaller constellation, but with each satellite being larger. One contender is 4piXIO – the 4p X-ray Imaging Observatory[57]. This mission would use coded aperture imaging[58], sometimes called "shadow mask imaging", to cover a huge field of view (some 45°×45°) with each telescope, yet return similar arcsecond level positions.

Exoplanets, planets orbiting stars other than our own Sun, have become one of the biggest and most exciting fields within astronomy over the past twenty years, and will be covered more comprehensively in the next section. Counter intuitively, high energy radiation is important for exoplanets both because the activity it tracks in the host star can be destructive to the planet's atmosphere, and because X-ray transits can measure the rate at which the atmosphere is being lost. The prospects for life on exoplanets are poor if these effects are large. Fortunately, advances in compact, lightweight, X-ray optics[59] now allow a search for the ionised atmospheres of planets around bright stars with a smallsat. Preliminary evidence from large X-ray observatories is that these are highly extended and may be being blown away[60]. Smallsats have the advantage that they can be dedicated to a single program, and so can observe eclipses for a modest number of stars many times in a year.

Observational opportunities in the UV and EUV (extreme ultraviolet) are scarce. Only NASA's Hubble and Neil Gehrels Swift observatories have a UV capability at present, and none have EUV capability. An observation of an unexpected UV flare from a Gravitational Wave event electromagnetic counterpart was a major inspiration behind the Israeli "ULTRASAT", a < 1 cu.m 250 kg mission[61] with a 50 cm mirror due for launch in 2022, and NASA's 12U "Gravitational-wave Ultraviolet Counterpart Imager" (GUCI) Mission[62]. Similarly, interest in the EUV is growing again, after years of quiet. The lack of EUV facilities is largely due to the roughly 300 light-year horizon in the EUV, due to absorption by interstellar hydrogen in the Milky Way, putting most astronomical objects out of reach. The discovery of exoplanets well within that horizon, and the realization that EUV radiation can be destructive to planet atmosphere, has made EUV observations of exoplanet host stars a priority. NASA has funded a study of one such smalsat mission, NExtUP, dedicated to this question, along with a larger - but still modest - small explorer Extreme-ultraviolet Stellar Characterization for Atmospheric Physics and Evolution (ESCAPE).

# Exoplanets

Exoplanets are still a young field, being driven by ideas and serendipity, so many discoveries can be made with modest facilities. The discovery of the first planet around a main sequence star[63], was awarded in 2019 with the Nobel prize for Physics. Interestingly, said discovery was made with a ~2m telescope at the Observatoire de Haute-Provence. Another key example is the more recent exceptional discovery[64] of a planetary system made of Earth- and sub-Earth-size planets around the ultracool dwarf TRAPPIST-1 enabled by two 60 cm telescopes at the La Silla Observatory in Chile and the Observatoire de l'Oukaïmeden in Morocco.

The Canadian *Microvariability and Oscillations of Stars telescope* (MOST[65]) was the first microsatellite in orbit. Launched in 2003, its primary goal was to monitor variations in star light over long periods of time (up to 2 months). MOST made major astronomical discoveries, e.g. that the star Procyon does not oscillate, contradicting previous observations made from the ground. MOST also showed for the first time that some hot giant planets orbiting very close to their star have a very low albedo (< 0.1 for HD209458b) ruling out the presence of high-reflective clouds in their atmospheres. MOST's cost was less than $ 6 million, including design, construction, launch and operations.

CNES-CoRoT (*Convection, Rotations et Transits planétaire*), launched in 2006, was the first space mission dedicated to detect exoplanets. The spacecraft, based on a CNES-Alcatel PROTEUS low-Earth orbit recurrent platform, was equipped with a 27 cm-diameter telescope. However, several sub-systems were upgraded, in particular to reach the stringent satellite pointing stability requirement of 0.5 arcseconds rms[66]. The mission's cost amounted to €170 million. Among the notable discoveries was CoRoT-7b, which was the first exoplanet shown to have a rock-dominated interior.

The search for new worlds in our galaxy in the past couple of decades has been highly successful and the prospects for the incoming decade are even brighter. Over 4300 exoplanets are known today, a large majority of which were discovered by the NASA Kepler mission, launched in 2009 in a trailer orbit[67]. The project's life-cycle cost, including 3.5 years of operations and excluding launch, was approximately $600 million. Kepler provided for the first time the demographics of planets in our galaxy. In the sample identified by Kepler there is little resemblance with the Solar System paradigm. For instance, planets with radii between that of the Earth and Neptune are by far the most common (e.g.[68]), while our Solar System has no example of these intermediate planets. These unexpected and intriguing results have prompted a great deal of effort in the community to increase the number of known extrasolar planets and

overcome the limits of the current incomplete sample, which still show biases imposed by the detection techniques adopted.

Focusing on current space missions to detect new exoplanets, NASA's *TESS* (Transiting Exoplanet Survey Satellite[69]) mission was launched in 2018 to an inclined, elliptical, lunar-resonant orbit around the Earth to survey the whole sky to search for exoplanets orbiting nearby stars. The ESA CHEOPS mission was launched in December 2019 with the scientific goal to search for transits of exoplanets around stars already known to host planets[70]. CHEOPS operates at 700 km in a Sun-synchronous orbit. Both TESS and CHEOPS are relatively small satellites optimised to observe bright targets, which are the most favourable for follow-up observations of their masses and atmospheric characterisation. Their costs are short of $300 million for TESS and about €150 million for CHEOPS.

Further into the future (2026), the ESA PLATO mission will aim at finding small planets transiting in the habitable-zone of solar-type stars[71], a goal that requires extremey high pointing stability and photometric precision and justifies the choice of a medium-size mission operating in L2 (M3 ESA budget €650 million plus payload costs).

Meanwhile, a new generation of cubesats is generating important and exciting science in preparation or in parallel to more ambitious missions. The *ASTERIA* mission[72] is an example. *ASTERIA* is a 6U cubesat capable of pointing with sub-arcsecond stability. *ASTERIA* measured the transit depth of a known planet, 55 Cnc e, only twice the size of the Earth and orbiting a very bright star (Fig. 2). As of February 2018, ASTERIA had met its primary mission requirements by demonstrating pointing stability better than 0.5 arcseconds RMS over 20 minutes and pointing repeatability of 1 milliarcsecond RMS from orbit-to-orbit. The mission also demonstrated thermal stability of +/-0.01 K as measured at a single point on the focal plane. ASTERIA serves as a demonstration of what a large constellation of such satellites could do.

What are these planets actually like, and –most importantly – why are they as they are? In the next decade, emphasis in the field of exo-planetary science will expand towards the understanding of the nature of the exo-planetary bodies and their formation and evolutionary history. This goal is achievable through the remote sensing observation of a large sample of exoplanetary atmospheres.

In the past ten years, we have learned how to obtain the first spectra of exoplanetary atmospheres. With the high stability of Spitzer, Hubble, and large ground-based telescopes, the spectra of close-in planets around bright stars were obtained. Key molecular, atomic and ionic species have been detected and the structure and stability of those atmospheres have been sounded. About 50 gaseous planets' atmospheres and the atmospheres of a few but very interesting small planets (e.g.[73,74]) have been observed so far. Despite these successes, current exo-atmospheric observations are still sparse and give only an incomplete picture of the planets studied. Additionally, flagship space observatories are heavily oversubscribed: the exoplanet oversubscription was a factor of 12 for Spitzer, while the Hubble Space Telescope is typically a factor 6.5 oversubscribed for exoplanets.

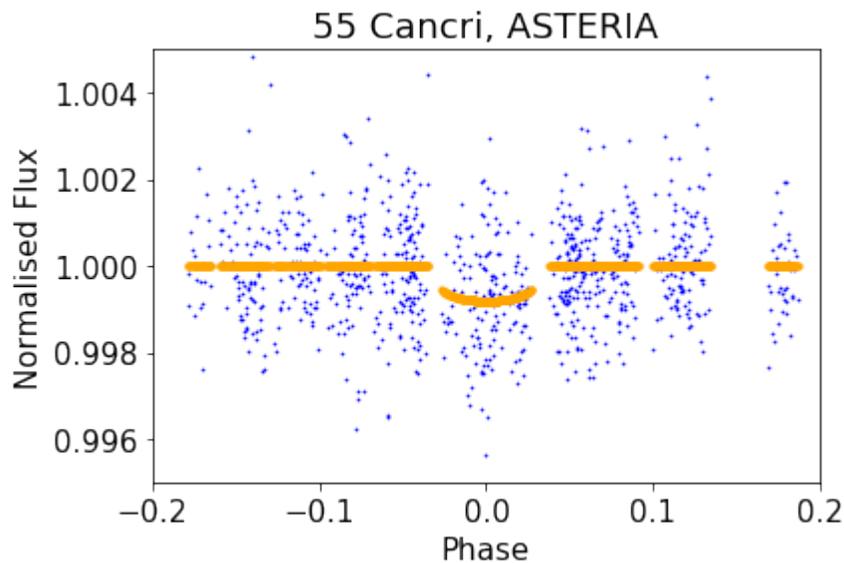

*Figure 2 Exoplanet transit observations by the ASTERIA mission, from [72]. Orbital phase-folded data points are shown in blue, and the model transit is shown in orange. Note the dip around phase zero, where the exoplanet 55 Cancri e passes in front of its star, 55 Cancri. The fluxes have been corrected for gain variations and normalised to an uneclipsed flux of unity.*

In the next decade the launch of JWST[26] will address many of the current instrumental limitations and its sensitivity will offer the opportunity to increase the number of small planets analysed, including the ones in the habitable-zone of cool stars, such as some of the TRAPPIST-1 planets. The ESA Ariel medium-size mission (M4 ESA budget €450 million plus payload costs), to be launched in L2 in 2028, has been designed to provide high-quality, optical to infrared spectra for a large sample (~ 1000) of diverse exoplanetary atmospheres[75].

While being JWST and Ariel transformational for exoplanet characterisation, they will not cover e.g. the UV spectral range. Their observing time will be precious: JWST is expected to be even more oversubscribed than HST and Spitzer[76]. There is therefore a clear niche for small missions before or in parallel the launch of JWST and Ariel to maximise their science return and complete the big picture. For instance, CUTE is a 6U CubeSat designed to observe in the near-ultraviolet (NUV) ~12 giant planets where atmospheric escape and interactions with the host star have the largest observable effects[77]. BSSL-Twinkle is a small satellite designed to provide spectroscopic measurements in the optical and infrared of exoplanets around very bright stars and Solar System objects. Twinkle will operate at 700 km in a Sun-synchronous orbit[78]; the data recorded by Twinkle will be delivered to scientists through a membership access model similar to the one implemented today for many ground telescopes.

# Solar System Applications

There are also astrobiological applications for small satellite technologies. The potential science questions here are both biological and chemical. The biological questions include, for example: can microbes survive transit from Mars to Earth, and survive re-entry? The context of this question is panspermia, i.e. the proposition that life originally started elsewhere such as on Mars, but then propagated to Earth via hypervelocity impacts[79]. Little enough is known about how life started to be able to rule this out. The chemical questions include, for example: how do complex pre-biotic molecules (e.g.[80]) form in space? Here the reactions are typically in the context of cryogenic surface chemistry on dust grains, in ultra-hard vacuums (e.g.[81]). Broadly speaking there is little gas-phase chemistry in astronomy because the interstellar and interplanetary media are so sparse, but if atoms or molecules are adsorbed onto a dust grain they can perform random walks across the surface and potentially interact. It is not easy to create laboratory analogues for space environments in terrestrial laboratories (e.g.[82]), so there are several experiments to test the survivability of bacteria in the harsh radiation environment of space, and the temperature variations and the ultra-hard vacuum, for example the EXPOSE experiment[83] that is mounted on the International Space Station (ISS). However, the ISS is not an ideal location for this experiment, because of the environmental pollutants from rocket propellant and other local emission sources. Furthermore, the radiation environment of the ISS is by design more benign than some experiments seek to investigate. There are therefore several cubesat experiments for astrobiology. If the satellite is placed in a polar orbit or similar, it can pass through the Van Allen belts and obtain a higher radiation dose. For example, the NASA O/OREOS 3U cubesat achieved 15 times the ISS exposure. There is still a large astrobiological parameter space still to explore. Aspirations include in-situ Raman or near-infrared spectroscopy, removing the need to recover the satellite, and there are aspirations to explore the effects on metabolically or reproductively active organisms (e.g.[84]).

Microsatellite constellations may also be an appropriate technology for monitoring the Earth's geomagnetic environment. The three LEO satellites of the SWARM mission[85], launched in 2013, monitor the strength and structure of the Earth's magnetic field and its effects on global positioning systems. While microsatellite swarms would have reduced capacity for instrumentation than the SWARM mission, they could still increase the instantaneous three-dimensional coverage of the geomagnetic environment.

Another use case for monitoring the Earth's environment is Near-Earth Asteroids (NEOs). These are often located at low solar elongation angles, so are affected by scattered sunlight in ground-based observations during daytime or twilight. The Near Earth Object Surveillance Satellite (NEOSSat), launched in 2013, has therefore been monitoring for NEOs from a microsatellite platform[86]; a microsatellite constellation could increase instantaneous areal coverage and improve completeness, as well as improving temporal coverage for light curves (variability due to rotation).

Finally, solar observations are not photon-starved so can be well-suited to small satellite platforms. For example, the *Miniature X-Ray Solar Spectrometer* (MinXSS) CubeSats[87] aim to monitor the solar X-ray flux (0.5-30 keV), infer coronal element abundances and determine the influence on the Earth's ionosphere and thermosphere.

# Technological challenges

Small satellites are often used to increase the Technology Readiness Level (TRL) of a satellite component by demonstrating capability in an operational environment, but some of the applications in this article call for dedicated technological development.

Low-Earth Orbit environment is an extremely challenging one. LEO commercial platforms usually do not require stray-light control or pointing stability at the level of an astrophysical mission. They can be modified by careful studies of additional payload elements (Earth-shield, baffle, fine-guidance-system, etc..) but clearly at an additional cost. For instance, in the case of CoRoT, at platform level the pointing was provided by star trackers, inertial wheels, magneto-torquers, and gyrometers, giving an angular stability of 16 arc seconds[66]. Since CoRoT's requirement was about 0.5 arcsec rms, to reach this value it was necessary to include ecartometry computations in the control loop based on the position of two stars on the payload focal plane in the control loop ecartometry.

The performances demonstrated in 2018 by CubeSat ASTERIA in terms of pointing stability and repeatability are very encouraging.

LEO satellite presents a one-sided view to a 220K black-body with emissivity and temperature variations as the orbit is completed which is not favourable to a thermally controlled environment or for potential telescope side-lobes introducing additional unwanted radiation as well as that originating from the source in question. Satellite thermal control is well-demonstrated on large missions with high TRLs, but miniaturisation is an additional obstacle for small satellites, and the cooling systems also need to satisfy the mass and power budgets of a small satellite. Passive systems (i.e. those requiring no power) such as thermal coatings and sunshades require less additional development than active (i.e. powered) systems such as cryogenic coolers, or cryocoolers. Miniaturised active cooling is needed for observations extending into the thermal infrared (longward of ~2$\mu$m), and is under development[88]. For example, the Cryocube-1 mission[89] was deployed from the ISS in February 2020. It is intended to reach 100K through passive cooling alone, and will perform cryogenic fluid management tests for future miniature active systems.

Deployable optics are another area where technological development would assist small satellite applications for astronomy. The size of the telescope primary mirror determines not just the angular resolution, but also the total light-gathering ability and hence the system signal-to-noise for any observation. There are therefore strong scientific drivers for maximising the primary mirror size. Launch vehicle volume constraints have driven the JWST design[26] to have unfolding segmented mirrors, while small satellites have their own volume constraints that also suggest deployable optics, particularly to the theoretical diffraction limit if possible. Several design studies have been conducted[90], but to date only the PRISM mission[91] is known to have made an in-flight demonstration, and then only to deploy lenses on an extending optical bench. Further work is merited.

Timing to microsecond accuracy is also critical for the high energy use cases discussed in this article, and here the situation is slightly more developed. For example, the NASA Deep Space Atomic Clock[92], launched in 2019, aims to maintain a timing precision of less than one microsecond per decade, in a 16kg instrument less than 30cm cubed.

# Outlook

As the cost and technical barriers to small-satellite astronomy are reduced, one possible over-arching consequence of increasing the numbers of PI-led missions is the impact on and from Open Science. Observatory missions typically have raw data proprietary lifetimes of 6-12 months, while PI missions and instruments sometimes have much more restrictive policies. Initiatives such as the European Open Science Cloud[93] aim to make all data FAIR (Findable, Accessible, Interoperable, Reusable), which in astronomy usually involves, among other things, the integration of products into the Virtual Observatory[6]. Meanwhile, data management and deposition plans are increasingly being required by funding bodies. Commissioning novel instrumentation can sometimes be a sound justification for extended proprietary lifetimes, but existing software, data standards and data repositories can still represent a significant cost saving to a small mission. The future small-satellite astronomy community could therefore do well to maintain the open data culture from observatories, not just for exploiting existing standards and repositories, but also for teams to be eligible for follow-on funding in the current climate of open data.

Commercial satellite swarms are a two-edged invention for astronomy. They are already known to pose many problems for ground-based and space astronomy, and industry-academia discussions are underway on mitigations; on the other hand the technology in itself has many other promising applications. Astronomy has a long track record of major discoveries following the opening of new observational parameter spaces (e.g.[94]), in turn driven by technological advances. This review shows that microsatellites and microsatellite constellations certainly have the capacity to open up new parameter spaces. The ingenuity of scientists virtually guarantees that there are many more potential astronomical applications to come.

# Acknowledgements

We thank the participants of the symposium on small satellites in London, November 2019 for many thoughtful contributions and for inspiring some of this Perspective. S.S. also thanks Karen Olsson-Francis and Hugh Dickinson for helpful discussions, the UK Space Agency for support under grant ST/T003502/1, and the ESCAPE project; ESCAPE - The European Science Cluster of Astronomy & Particle Physics ESFRI Research Infrastructures has received funding from the European Union's Horizon 2020 research and innovation programme under Grant Agreement no. 824064.